# Detecting Antihydrogen: The Challenges and the Applications[*]

Makoto C. Fujiwara

*TRIUMF, Canadian National Laboratory for Particle and Nuclear Physics*
*4004 Wesbrook Mall, Vancouver, British Columbia, V6T 2A3, Canada*

**Abstract.** ATHENA's first detection of cold antihydrogen atoms relied on their annihilation signatures in a sophisticated particle detector. We will review the features of the ATHENA detector and its applications in trap physics. The detector for a new experiment ALPHA will have considerable challenges due to increased material thickness in the trap apparatus as well as field non-uniformity. Our studies indicate that annihilation vertex imaging should be still possible despite these challenges. An alternative method for trapped antihydrogen, via electron impact ionization, will be also discussed.

**Keywords:** Antihydrogen, Fundamental Symmetries, Si vertex detectors.
**PACS:** 52.27.Jt, 36.10.-k, 39.10.+j

## INTRODUCTION

A long term goal of antihydrogen research is precision tests of CPT and other fundamental symmetries via spectroscopy comparisons of hydrogen and its antimatter counter part, antihydrogen. The ATHENA experiment, located at CERN, achieved a major milestone by demonstrating production of cold antihydrogen atoms [1]. The ATHENA data taking is now completed, and a new experiment is proposed to continue its effort. ALPHA, Antihydrogen Laser Physics Apparatus, will aim to trap cold antihydrogen in a magnetic trap [2]. Stable trapping of antihydrogen will likely open up various possibilities in fundamental physics with cold anti-atoms. This article reviews the design issues of the ALPHA antihydrogen vertex detector.

## ANTIHYDORGEN DETECTOR

The capability to detect the vertex position of antihydrogen annihilations was an important feature of ATHENA not only for the identification of cold antihydrogen production [1], but also in the developments leading to the first production [3, 4]. Having real time images of antiproton losses was essential as a diagnostic when optimizing the trapped particle manipulation in a nested Penning trap.

In order to trap neutral antihydrogen atoms in ALPHA, we plan use an octupole magnetic field configuration for radial confinement, and a mirror field configuration

---

for axial confinement. As we enter the unexplored regime of neutral anti-atom trapping, we believe it is crucial to retain the vertex imaging as a diagnosis tool. There exists very little experimental information on the behaviour of trapped particles in multipolar, and mirror magnetic fields. We will likely encounter unexpected new effects. Given the situation, the importance of real time imaging, together with other plasma diagnosis techniques we developed for ATHENA [5, 6], cannot be overstated.

We note further that vertex detection capability will play an important role in the future phases of ALPHA. Even after stable trapping of neutral antihydrogen atoms is achieved, performing spectroscopy measurements with few atoms will be extremely challenging, given the expected low signal rates. A clean vertex determination will help demonstrate laser transitions of anti-atoms, e.g. via resonant photo-ionization, and discriminate against possible sources of backgrounds.

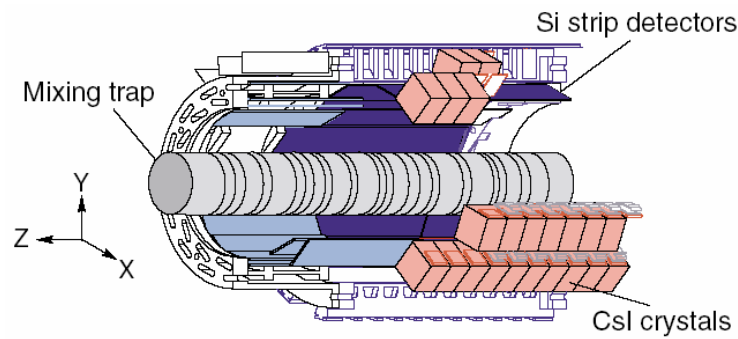

**FIGURE 1.** Schematic view of ATHENA antihydrogen detector [3, 4]. Two layers of double-sided Si strip detector and 192 CsI crystals surround the trap region. A 1.65 mm thick copper vacuum wall, between the trap and the detector, is not shown in the figure.

## THE CASE FOR VERTEX DETECTION IN ALPHA

In order to illustrate the power of the vertex detection, we will give in this section some examples from our experiences in ATHENA. Figure 2, taken from our first report [1], shows the difference in the vertex distributions between production data (cold mixing) and the control data (hot mixing). This provided an important piece of evidence for establishing the first production of cold antihydrogen atoms. Subsequent analyses using simulated vertex distributions showed that about 65% of annihilation events in Fig 2 (a) are due to antihydrogen and the rest due to background [7].

It should be stressed that detecting only antiproton annihilations, for example by external scintillators, is not sufficient evidence on its own for antihydrogen production, because of the existence of particle loss processes in the trap. Figure 3 further illustrates the importance of spatially sensitive detection. The axial vertex distributions are compared between the standard cold mixing and the "mixing" without positrons. A clear difference is observed. Suppression or inhibition of antihydrogen production due to causes such as plasma instabilities, vacuum deteriorations, or electrodes malfunctions, are sometimes otherwise difficult to identify, but they would show up immediately in the vertex distributions allowing rapid diagnosis of the system. An incomplete removal of the electrons from the trap (which is used to cool antiprotons in the first step of the mixing cycle) results in the vertex pattern similar to Fig. 3 (b) indicating suppressed antihydrogen production.

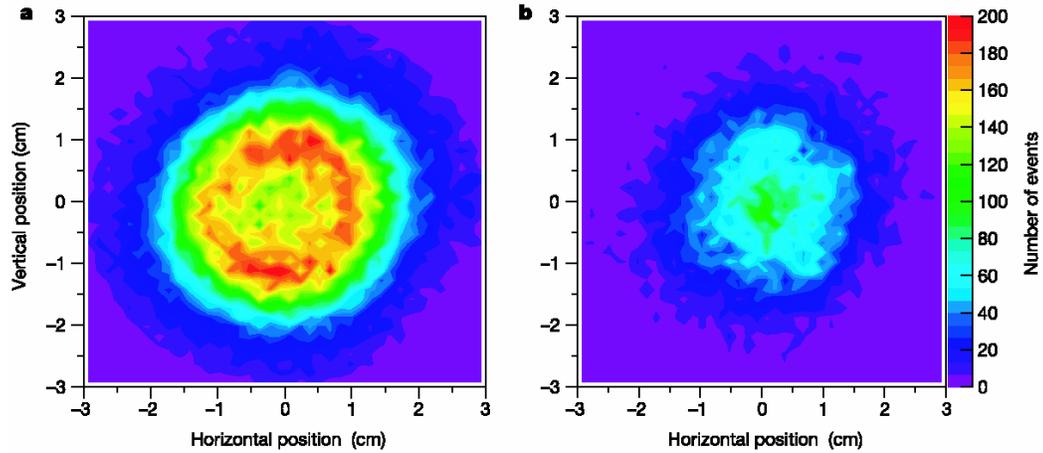

**FIGURE 2.** Schematic view of Fig. 2: X-Y distribution of antiproton annihilation vertices for (a) cold mixing and (b) hot mixing, obtained with the ATHENA detector. Both plots are normalized to the same number of mixing cycle. Enhanced annihilations on the trapped electrodes (inner radius 1.25 cm) in Fig (a) indicates antihydrogen production, whereas annihilations of the central part in Fig (b) represents antiproton annihilations on residual gas or ions.

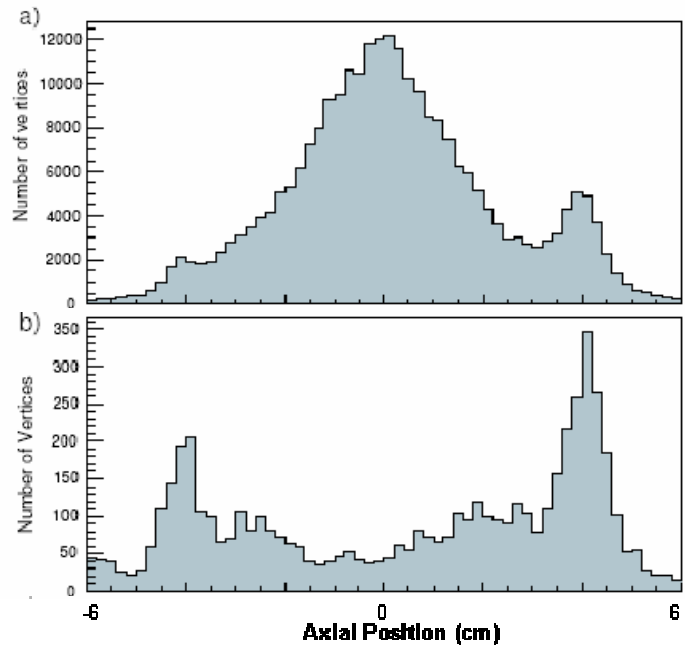

**FIGURE 3.** Axial (i.e. z) distribution of annihilation vertex for (a) standard cold mixing and (b) the data without positrons. Figure taken from [8].

While these observations attracted some interests on their own from the trap physics point of view, they have important implications for antihydrogen detection. Recall the radially symmetric distribution of annihilation at the trap wall for antihydrogen (Fig 2 a). The observation that *charged* particle loss at the wall results in hot spots, while the neutral antihydrogen atoms annihilates symmetrically, provides a new and effective signature of antihydrogen identification. An advantage is that we do not need to rely on the 511 keV detection, which are difficult due to its low efficiency and large background. In fact, in the recent runs of ATHENA, the presence or absence of the hot spots has been one of the most valuable diagnostic of our antihydrogen production processes.

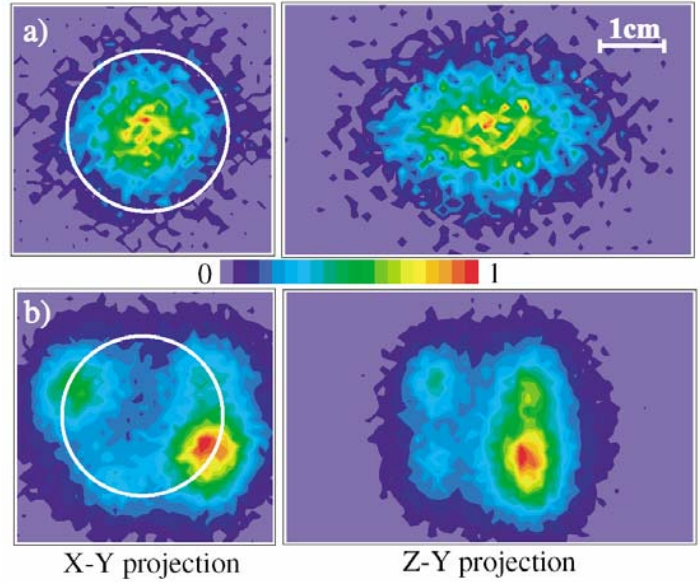

**FIGURE 4.** Annihilation images of trapped antiprotons in a harmonic Penning trap with (a) high residual gas density, and (b) low residual gas density. Both data are without positions. The electrode position is shown with a circle [4].

# DESIGN CONSIDERATIONS

The overall design of the ALPHA experiment will be driven by the requirements for neutral atom trapping. This implies, for the Si detector, that there will be a substantial amount of superconducting magnet materials between the trap and the detector. Our design goal is to retain the vertex position resolution similar to ATHENA of 4 to 5 mm ($\sigma$) as illustrated in the previous section.

## Modifications and Improvements over the ATHENA Detector

The ATHENA detector has worked well for the most part as discussed above, and we plant to adopt its basic design features. However, several modifications and improvements are foreseen.

The ATHENA detector did not allow the determination of the curvature of charged particle tracks in a 3T magnetic field, since it only had two double-sided layers. Hence the charged tracks were approximated with straight lines, and this was a dominating factor in the vertex resolution of 4-5 mm (see below). We plan to improve the tracking in the ALPHA vertex detector by having three or more Si layers. Gaseous detectors such as a multi-layered drift chamber or a time projection chamber would be an alternative possibility. The improved determination of the charged tracks will partly compensate the resolution loss due to the increased materials between the trap and the detector which are unavoidable.

511 keV gamma detection will be no longer essential in antihydrogen identification for the new generations of experiments such as ALPHA, as shown in Ref. [4, 7]. However, we will have a limited number of gamma detectors inside our magnet for the diagnosis of trapped positrons. Transmission of 511 keV gammas through a 1 cm thick Cu is of order of 25%, allowing sufficient count rates for the diagnosis of trapped cold positron plasma [9].

We envision operating the ALPHA detector at a temperature sufficiently higher than the ATHENA one, which were kept at 140K. Some of the problems we encountered in the ATHENA detector, in particular, deterioration of triggering capability and a steady increase in inactive modules, may be attributed to low temperature operations and repeated thermal cycling. By operating at a higher temperature, we hope to avoid these risks.

## Multiple Scattering

A very rough estimate of multiple scattering in the ALPHA neutral trap magnet can be obtained from the approximate Moliere formula [10]:

$$\theta_0 = \frac{13.6 \text{ MeV}}{\beta c p} z \sqrt{x/X_0} [1 + 0.038 \ln(x/X_0)] \qquad (1)$$

where $\theta_0$ is an rms scattering angle (projected on a plane), $p$, $\beta c$, and $z$ are the momentum, velocity, and charge number of incident particles, and $x/X_0$ is the thickness of the material in radiation length. In our case, $z=1$, $\beta \sim 1$, and the average pion momentum $p \sim 300$ MeV/$c$. Our neutral trap magnet will be made of a superconducting alloy of NbTi/Cu. If we assume a Cu equivalent thickness of 1 cm for the scattering medium, we have $\theta_0 \sim 40$ mrad. For an averaged distance between the vertex and scattering material (i.e. the lever arm) of 4 cm, this corresponds to $\sim 2$ mm error in the measured track position at the vertex. This rough estimate indicates that the contribution to vertex resolution from the increased material thickness is comparable or smaller than the total resolution of the present ATHENA, the latter being dominated by the unmeasured track curvature. Therefore, if sufficiently accurate determination of the track curvature can be achieved by three or more layers of Si, the increased materials in ALPHA will be manageable in terms of vertex detection.

Detailed GEANT3 simulations have been performed to better estimate the effect of multiple scattering. Vertex resolutions, defined here as the rms distance between the true vertex and the reconstructed vertex, were calculated for different thicknesses of the material between the trap and the detector. Antiprotons are annihilated on the trap wall and pions are generated with realistic phase space and branching ratios. The straight line approximation was used for reconstruction routine. Table 1 summarizes the results of this simulation study.

**TABALE 1.** Simulated vertex resolutions ($\sigma$) for various material thicknesses

| Material Thickness Cu equivalent (mm) | Solenoid Filed | Resolution x (mm) | Resolution z (mm) | Comments |
|---|---|---|---|---|
| 1.65 | 3 T | 4.3 | 1.6 | ATHENA case |
| 1.65 | 0 T | 1.3 | 1.7 | ATHENA- without B field |
| 10.0 | 3 T | 5.0 | 3.0 | ALPHA - two Si layers |
| 10.0 | 0 T | 2.8 | 2.8 | ALPHA- known track curvature |

The first row is similar to the ATHENA case, where the Cu vacuum wall is 1.65 mm. An X resolution, much worse compared to Z, reflects the effect of magnetic field. This is confirmed in the second row, where the B field is turned off, and the X resolution is substantially improved. The remaining resolution in Row 2 is due primarily to multiple scattering. This comparison of the two cases indicates that ATHENA's 4 mm vertex resolution is dominated by the unmeasured track curvatures.

The third and forth rows are the simulations with increased scattering material. Instead of 1.65 mm, a 1 cm thick Cu, which is similar to the thickness of superconducting magnet material in ALPHA, is used in the calculations. With a 3 T solenoid field on (Row 3), the resolution is worsened compared to the standard ATHENA case (Row 1), due to the increased multiple scattering as expected. Recall that because of the straight line fits in the track reconstruction, the track curvature is unmeasured in this case. This level of vertex resolution is expected when only two Si layers are used in ALPHA. Row 4, with no B filed, isolates the effect of multiple scattering as in Column 2. The X plane resolution of 2.8 mm is in rough agreement with the estimate above using the Moliere formula. This value in Column 4 indicates that, even with a thick magnet, if the reconstruction error due to track curvature can be made negligible (e.g. by having three Si layers), the vertex resolution of the order of 3 mm can be achieved. This is well within our design goal.

The calculated distributions of the difference between true vertices and the reconstructed vertices, X(MC)-X(reconst), are graphically compared for the cases without magnetic field (Fig. 5, left), and with the 3T solenoid field (Fig 5, right).

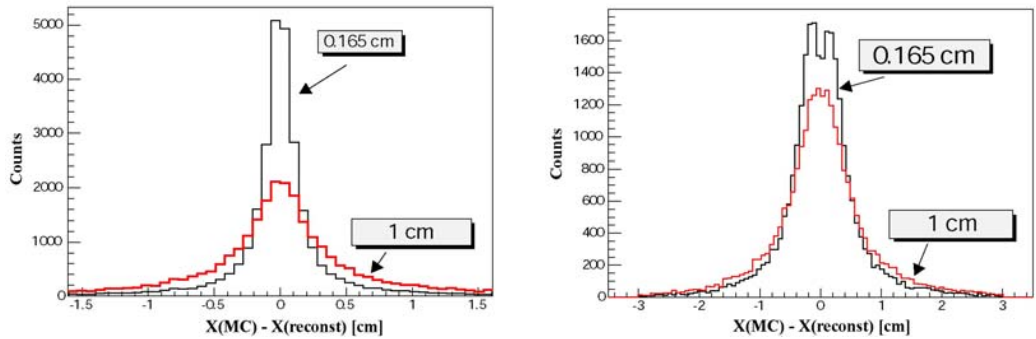

**FIGURE 5.** Calculated difference between the true vertices and reconstructed vertices for two different material thicknesses (as indicated in square boxes). Left: with out magnetic field, Right: with 3T solenoid field.

Note that magnitude of multiple scattering scales only as square root of the material thickness (see Equation 1), hence our conclusion here is not overly sensitive to the exact thickness of the magnet materials in the ALPHA apparatus.

## Other Effects

The effect of the multipolar field (as opposed to solenoid) on charged particle trajectories is expected to be small compared to the multiple scattering discussed above, and can in principle be corrected, given the charge and momentum of pions, and the magnetic field distribution. The axial field variation due to the mirror magnet could range from 1T to 3T, and its effect on the vertex resolution may be larger than the radial multipole field. A quantitative GEANT study is under way to study these effects.

Increased thickness of magnet material would result in a greater probability for the conversion of high energy gammas from neutral pion decays. The track pattern recognition routine in the off-line analysis software may need to be improved, should the increased multiplicity of charged tracks become problematic. Note that conversion events will likely have well-defined topology, since they mostly occurs at the neutral trap magnet at a fixed radius.

## TRIGGER AND DATA ACQUISITION

Improvements are foreseen for trigger and data acquisition system for the ALPHA detector, compared to ATHENA. A basic trigger scheme is illustrated in Figs. 6. A good feature of the IDEAS VA-TA readout chips used in ATHENA was their self-triggering capability. In our analysis [7], we showed that this trigger signal can be used as a proxy for antihydrogen signal in many cases, and some important physics results were obtained using this level 0 trigger signal [9,10]. We plan to retain this capability in the ALPHA. Higher level triggers (Fig. 6) can apply various cuts such as the multiplicity and the event topology as well as the trap conditions.

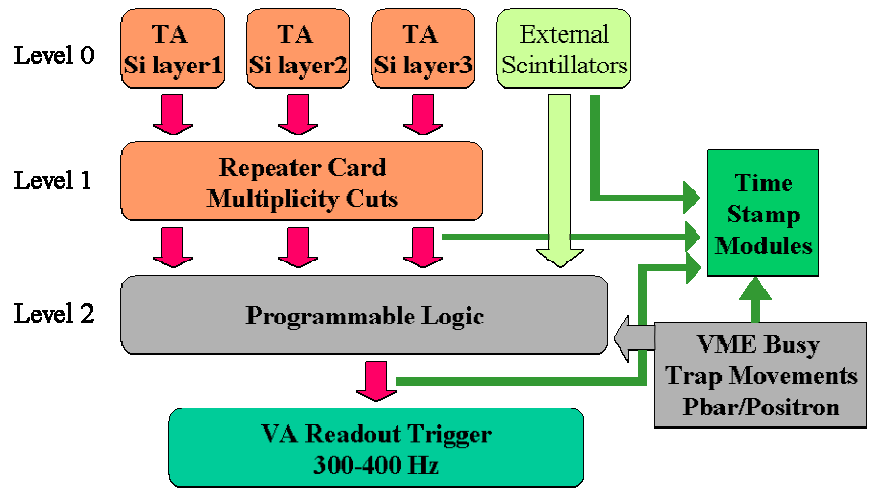

**FIGURE 6.** Vertex detector trigger scheme

Trigger events at each level, external annihilation scintillator signals, together with trap and other slow control activities, are time-stamped via multi-scalars which are synchronized to a 10 MHz atomic clock. Deadtime-less operation is possible for indefinite duration.

## ALIGNMENT

Relative mechanical alignment of each layer of Si detector with respect to one another will be determined using cosmic rays when the magnetic field is turned off. Alignment of the detector with respect the trap is somewhat more difficult, and we will use a method developed for the ATHENA detector [13]. Figures 8 and 9 show examples of such measurements. By moving the antiproton trap well, and measuring the annihilation positions, one can obtain the correlations between trap well positions and measured annihilation positions. The detector $z$-position with respect to the trap is thus determined within 1 mm accuracy, a task which is otherwise difficult to achieve in our setup.

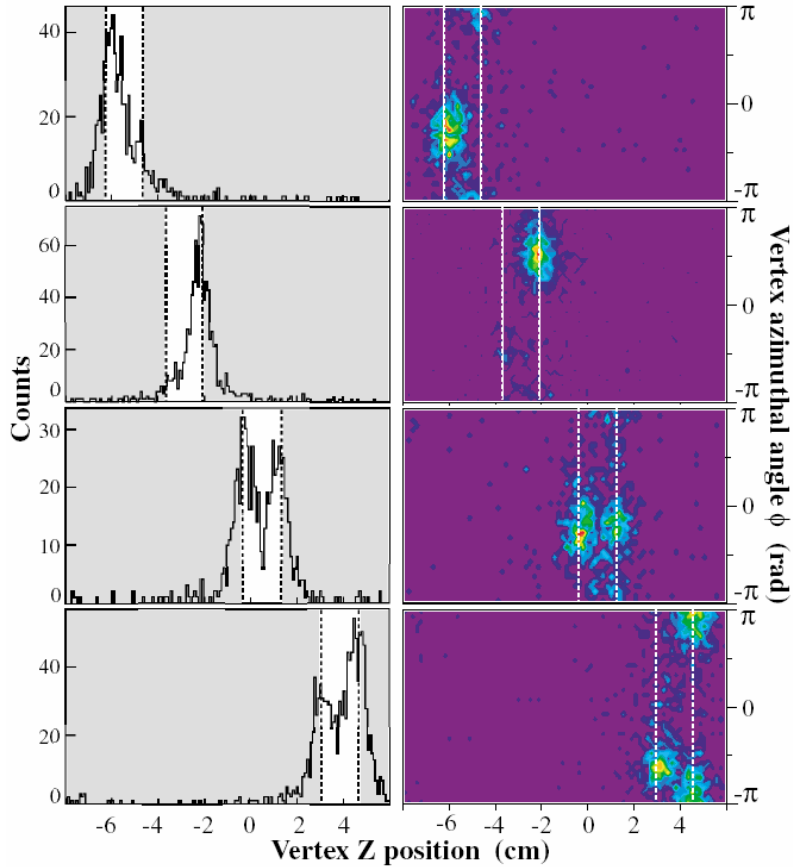

**FIGURE 8.** The projection of the antiproton annihilation distribution on the z axis (left column) and on the z-phi plane (right column) for four different confinement setup. The trap well positions are indicated by the unshaded regions, and the positions of the electrodes are depicted with dashed lines.

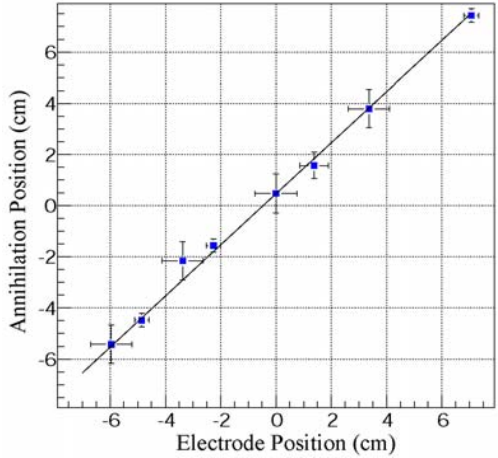

**FIGURE 9.** Correlation between the trap well position and the measured annihilation positions.

## ALTERNATIVE DETECTION SCHEME

A straightforward approach for demonstrating antihydrogen trapping would be to ramp down one of the trapping magnets, and observing the annihilations of the antihydrogen atoms leaking out of the trap. A technical challenge of this method may be rapid ramping of the superconducting magnet, in a time scale shorter than the trap lifetime. As well, if the ramping is slow, the annihilation signals have to compete with cosmic and other backgrounds. Here we consider another method using electron impact ionization.

### Electron Impact Ionization of Antihydrogen Atoms

By bombarding a beam of the electrons onto the trapped antihydrogen, the anti-atoms can be ionized. The antiprotons from the antihydrogen ionization can be collected in an electric potential well. If the electric trapping potential is larger than antiproton recoil energy (which should be less than few eV), antiprotons will be "born trapped". After accumulating a certain number of ionized antiprotons in the collecting well, they can be dumped onto the degrader in a usual manner [14], providing an efficient, nearly background free detection. See Fig. 10 for a schematic drawing.

### Detection Efficiency

The efficiency of this detection method can be estimated by assuming that the cross section is similar to that for electron impact ionization of hydrogen atoms. The latter can be found, e.g. in Ref. [15]. The rate for ionization (1/s) can be written: $R = \sigma_i \times n \times L \times flux$, where $\sigma_i$ (cm$^2$) is the electron impact ionization cross section, $n$ (cm$^{-3}$) the trapped antihydrogen number density, $L$ (cm) the length of antihydrogen cloud, and *flux* (1/s) is the flux of electron beam interacting with the antihydrogen cloud. Using the cross section for hydrogen of ~ $10^{-16}$ cm$^2$ (at ~50 eV), and assuming 1000 antihydrogen is trapped in a 1 cm$^3$ volume (i.e. $nL$ ~ 1000 cm$^{-2}$), a 1 mA electron beam interacting with the antihydrogen would give 1000 Hz ionization rate, i.e. all the antihydrogen are ionized in 1 sec. We note that we can vary the antiproton catching potential shape to explore the special distribution of trapped antihydrogen cloud. A potential difficulty is the transport of the electron beam through the multipole magnetic field. Detailed studies are in progress to determine the feasibility of this method. We note that realistic theoretical calculations for the cross section for electron impact antihydrogen ionization will be helpful.

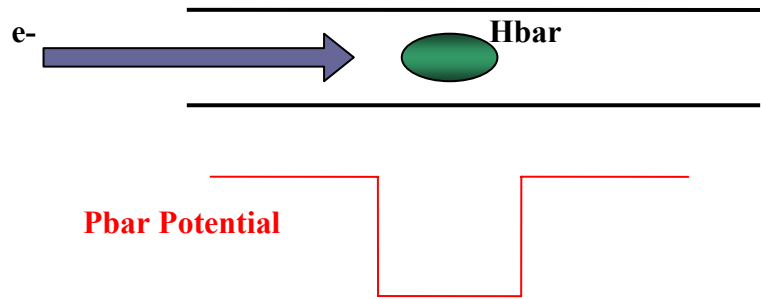

**FIGURE 10.** A schematic for electron impact ionization of trapped antihydrogen atoms.

## SUMMARY AND OUTLOOK

We have reviewed the features of the ATHENA vertex imaging detector, and the design challenges for a new detector for the ALPHA experiment. We have shown that the difficulties with unusually large amount of scattering materials could be overcome. Recently, this is confirmed via more detailed Monte Carlo simulations with three Si layers and helix fitting of the pion trajectories [16]. The final design of the detector is being worked out. An alternative method for trapped antihydrogen detection, based on electron impact ionization of antihydrogen, was also proposed in this article.

The ALPHA experiment aims to start its data taking at CERN AD in summer 2006. The vertex detector will be an unrivaled feature of ALPHA, which will allow detailed diagnostics of the trapped particles, as well as detection of cold antihydrogen atoms.

## ACKNOWLEDGMENTS

I would like to thank Professor Yasunori Yamazaki for the invitation to a very stimulating workshop. I also thank ATHENA and ALPHA collaborators for various discussions, and Professor Alberto Rotondi and Dr. Pablo Genova for valuable help with simulations. This work is supported in part by TRIUMF and by NSERC of Canada.